\documentclass[aip,rsi,reprint]{revtex4-1}

\usepackage{amssymb,amsfonts,amsmath}
\usepackage{graphicx} 
\usepackage{bm}        
\usepackage{color}
\usepackage{float}
\usepackage{subfigure}
\usepackage{physics}
\usepackage{bm}
\usepackage{epstopdf}

\begin{document}

\newcommand{\Caion}{Ca$^+$}
\newcommand{\Nion}{N$_2^+$}
\newcommand{\KkB}{K$\cdot$k$_\textrm{B}$}
\newcommand{\AX}{$A^2\Pi_u \leftarrow X^2\Sigma_g^+$}

\title{State-selective coherent motional excitation as a new approach for the manipulation, spectroscopy and state-to-state chemistry of single molecular ions}

\author{Ziv Meir}
\thanks{These authors contributed equally to this work}
\affiliation{Department of Chemistry, University of Basel, Klingelbergstrasse 80, Basel 4056, Switzerland.}
\author{Gregor Hegi}
\thanks{These authors contributed equally to this work}
\affiliation{Department of Chemistry, University of Basel, Klingelbergstrasse 80, Basel 4056, Switzerland.}
\author{Kaveh Najafian}
\thanks{These authors contributed equally to this work}
\affiliation{Department of Chemistry, University of Basel, Klingelbergstrasse 80, Basel 4056, Switzerland.}
\author{Mudit Sinhal}
\affiliation{Department of Chemistry, University of Basel, Klingelbergstrasse 80, Basel 4056, Switzerland.}
\author{Stefan Willitsch}
\affiliation{Department of Chemistry, University of Basel, Klingelbergstrasse 80, Basel 4056, Switzerland.}
\email{stefan.willitsch@unibas.ch}
\date{\today}

\begin{abstract}
We present theoretical and experimental progress towards a new approach for the precision spectroscopy, coherent manipulation and state-to-state chemistry of single isolated molecular ions in the gas phase. Our method consists of a molecular beam for creating packets of rotationally cold neutrals from which a single molecule is state-selectively ionized and trapped inside a radiofrequency ion trap. In addition to the molecular ion, a single co-trapped atomic ion is used to cool the molecular external degrees of freedom to the ground state of the trap and to detect the molecular state using state-selective coherent motional excitation from a modulated optical-dipole force acting on the molecule.
We present a detailed discussion and theoretical characterization of the present approach. We simulate the molecular signal experimentally using a single atomic ion indicating that different rovibronic molecular states can be resolved and individually detected with our method. 
The present approach for the coherent control and non-destructive detection of the quantum state of a single molecular ion opens up new perspectives for precision spectroscopies relevant for, e.g., tests of fundamental physical theories and the development of new types of clocks based on molecular vibrational transitions. It will also enable the observation and control of chemical reactions of single particles on the quantum level. 
While focusing on \Nion{} as a prototypical example in the present work, our method is applicable to a wide range of diatomic and polyatomic molecules.
\end{abstract}

\footnotetext{\ddag~These authors contributed equally to this work.}

\maketitle

\section{Introduction}

The spectroscopy of molecular ions has made substantial progress over the last decades \cite{willitsch11a}. Adiabatic cooling of ions in molecular beams \cite{bieske00a,duncan03a} and their cryogenic-buffer-gas \cite{gerlich95a,wester09a,rizzo09a,Campbell2015,asvany15a} or sympathetic cooling \cite{molhave00a,tong10a,willitsch12a,germann14a,Brown2018} in ion traps have enabled the study of the spectra of cold samples with unprecedented sensitivity and resolution. The majority of state-of-the-art experiments relies on sensitive action-spectroscopic techniques to detect spectroscopic excitations in molecular ions, for instance dissociation \cite{duncan03a,rizzo09a,Campbell2015}, chemical reactions \cite{schlemmer99a,tong10a,asvany15a} or the inhibition of cluster growth \cite{Chakrabarty2013}. All of these methods, except the last, entail a destruction of the ions for the determination of the final quantum state after photoexcitation and, therefore, necessitate a replenishment of the molecular sample in every experimental cycle. 

However, in the context of advanced applications such as precision spectroscopies \cite{schmidt05a,germann14a,kajita14a} and upcoming quantum technologies \cite{Blatt2008,haeffner08a,schmidt-kaler03a,mur-petit13a,leibfried12a,ding12a}, it is advantageous, and indeed in many cases necessary, to suppress ensemble averaging and work with small samples or even single particles which are not destroyed during the readout of their quantum states. Such methods are now routinely available for a range of atomic systems. Indeed, over the last few decades, there have been impressive advances in the coherent manipulation, control and readout of individual atoms. Atoms and atomic ions are readily trapped, cooled and coherently manipulated to form the basis of the most precise clocks \cite{Chou2010,huntemann16a,Bloom2014}, to perform quantum simulations \cite{Blatt2008} and to compute quantum algorithms \cite{Monz2016,Linke2017}. It is thus clearly desirable to adapt the techniques developed for the coherent control of atoms to molecules in order to enhance the scope of precision measurements on molecular systems \cite{carr09a,mur-petit12a,leibfried12a,ding12a}. In this context, one of the major challenges is the lack of optical cycling transitions in most molecules with the exception of a few specific systems with diagonal Franck-Condon factors \cite{shuman10a,Truppe2017,Anderegg2018}. These transitions are used in atomic systems for efficient state readout \cite{Myerson2008fix}. Thus, alternative techniques for the state readout of single molecules have to be implemented as is done in the present work.

Here, we present and discuss a quantum non-demolition approach \cite{Hume2007fix} for the detection of the quantum state of a single trapped molecular ion. The aim is the precise control and spectroscopy of a single molecule while conserving its internal state and chemical identity. The data collected in a single experiment is limited to the information obtained from a single particle and, therefore, does not benefit from the increased measurement statistics obtained from interrogating large ensembles. However, the duty cycle and therefore the measurement statistics of our experiment can be enhanced by several orders of magnitude compared to previous destructive readout schemes \cite{willitsch11a, willitsch12a} due to the non-destructive nature of our technique \cite{Hume2007fix}.

In our experiment, a single molecular ion is produced in a specific rovibronic state by photoionization from a molecular-beam source \cite{tong10a} and trapped in an ion trap. Co-trapped atomic ions are then used to translationally cool the molecular ion eventually forming a two-ion atom-molecule Coulomb crystal \cite{willitsch12a,willitsch17a}. The application of a state-dependent optical-dipole force (ODF) to the molecular ion modulated at a motional frequency of the ion string coherently excites motion in the two-ion crystal which is subsequently read out using the atomic ion with high sensitivity. A version of this scheme for molecules has originally been proposed in Ref. \cite{koelemeij07a} and experimentally demonstrated with atomic ions in Ref. \cite{hume11a}. In essence, this approach represents a quantum-logic spectroscopy (QLS)-type \cite{schmidt05a} state readout for molecules \cite{leibfried12a,ding12a,mur-petit12a} variants of which have recently also been realized with MgH$^+$ in Ref. \cite{wolf16a} and with CaH$^+$ in Ref. \cite{chou17a}. 

The flexible nature of the present experimental approach to study various types of molecules, together with their coherent control and state detection, makes it advantageous also in the realm of chemical studies. We envision experiments in which, e.g., the state of a single molecule and its chemical identity is monitored during collisions or reactions. This type of state-to-state chemistry experiments can shed light on inelastic and reactive processes on the single-molecule level \cite{hall11a,hall12a,willitsch17a} in a fashion similar to recent work investigating spin-conserving and non-conserving collisions in ultracold atom-ion reactions \cite{ratschbacher13a,sikorsky18fix,Furst2018,Sikorsky2018phasefix}.

In the present experiments, we focus on \Nion{} as the molecular ion of interest. \Nion{} features electric-dipole-forbidden transitions within the rovibrational manifold of its $X~^2\Sigma_g^+$ ground electronic state \cite{germann14a} which makes it an ideal candidate for a mid-IR frequency standard \cite{karr14afix,schiller14a,kajita15a} and for spectroscopically testing a possible time variation of the proton-to-electron mass ratio \cite{schiller05a,kajita14a}. However, the same features also hinder rovibrational state preparation by black-body-radiation-assisted optical pumping \cite{vogelius06a,wolf16a,chou17a} and require alternative state-preparation schemes as employed here \cite{tong10a,tong11a}. 



Achieving full quantum control over a single trapped molecular ion requires several abilities which are detailed in the following sections:
\begin{itemize}
    \item \textbf{Preparation} of the molecular ion in a specific initial internal quantum state (Sec. \ref{sec:prep})
    \item \textbf{Cooling} of its translational degrees of freedom to the quantum ground state (Sec. \ref{sec:cooling})
    \item \textbf{Detection} of its quantum state in a non-destructive manner with high fidelity (Sec. \ref{sec:detection})
    \item \textbf{Coherent manipulation} of an isolated sub-space of molecular quantum states (Sec. \ref{sec:coherent})
\end{itemize}


\section{Experimental system}\label{sec:exp_sys}

\begin{figure*} 
	\centering
	\includegraphics[width=\linewidth,trim={3cm 2cm 21cm 1cm},clip]{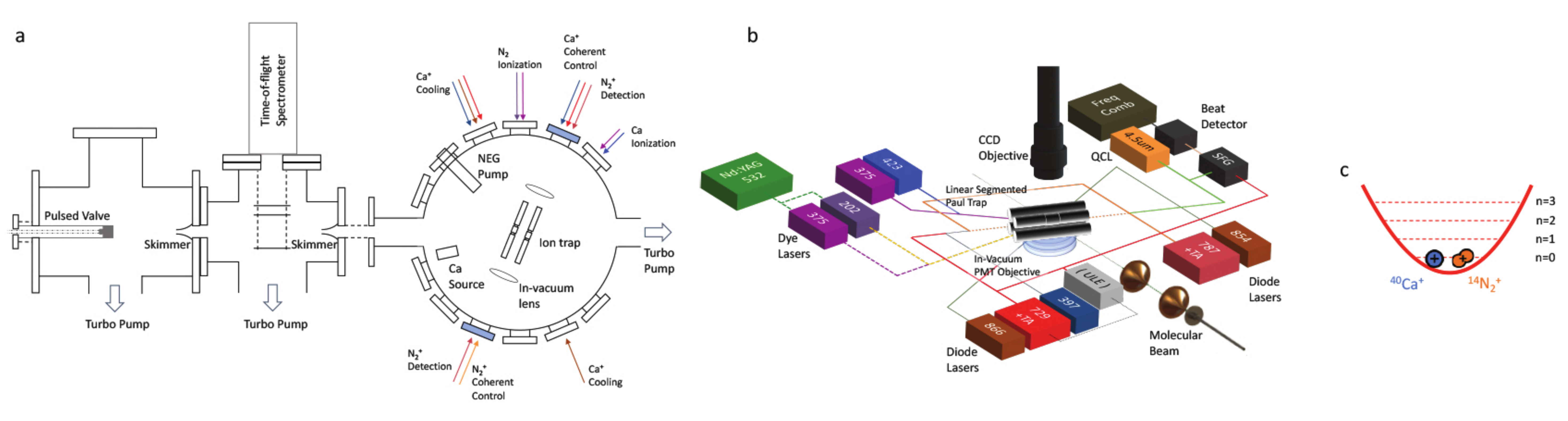}\\
	\caption{a) Schematic of the experimental setup. Two vacuum chambers used to create and direct a molecular beam are drawn from a side view while the science chamber housing the ion trap is drawn from a top view. Colored arrows represent different laser beams entering the science chamber used for ionization, cooling and coherent control of both the atomic and the molecular ion. b) Schematic of the laser and imaging systems used in the present experiment. Pulsed dye laser beams at 375 and 202 nm are used to ionize single N$_2$ molecules from the molecular beam (Sec. \ref{sec:prep}) while two continuous-wave diode laser beams at 423~nm and 375~nm are employed to ionize Ca atoms from an atomic beam. External-cavity diode lasers (ECDL) (397, 866 and 854 nm) and an amplified ECDL locked to an ultra-low expansion cavity (729 nm) are used for Doppler and ground-state cooling and coherent manipulation of the atomic ion (Sec. \ref{sec:cooling}). An amplified ECDL beam (787 nm) is split and then combined interferometrically to create a running one-dimensional optical lattice for the coherent motional excitation of the molecular ion (Sec. \ref{sec:detection}). A quantum-cascade laser (QCL) locked to a frequency comb will allow the coherent manipulation of the molecular ion in the infrared domain (Sec. \ref{sec:coherent}). Fluorescence of the atomic ion is imaged by a microscope objective coupled to an electron-multiplying charge-coupled-device (EMCCD) camera and an in-vacuum high-numerical-aperture lens system coupled to a photomultiplier tube (PMT). TA denotes tapered amplifiers for ECDLs. The color code for the laser beams is the same as in panel a. c) Illustration of a single-atom (\Caion) single-molecule (\Nion) Coulomb crystal (see Fig. \ref{fig:IonsPics}e for an experimental fluorescence image) cooled to the ground state of the harmonic trapping potential (see Sec. \ref{sec:cooling} and Fig. \ref{fig:GSCresults}). $n$ denotes the quantum number of the associated quantum harmonic oscillator.}
	\label{fig:exp}
\end{figure*}

Fig. \ref{fig:exp}a shows our apparatus which, to our knowledge, is the first to combine a molecular-beam setup with an ion-trap experiment capable of quantum logic. Other quantum-logic \cite{wolf16a,chou17a} and coherent ultracold-molecules experiments \cite{Ospelkaus2008} rely on chemical reactions to create the molecular species. Here, we use photoionization from a molecular beam as a general approach capable of producing a large variety of diatomic and polyatomic molecular ions. 

The setup is divided into three vacuum stages with decreasing pressures from high vacuum (5$\cdot$10$^{-6}$ mbar) to ultra-high vacuum (2$\cdot$10$^{-10}$ mbar). In the first chamber, we use a pulsed valve to create a molecular beam of internally cold molecules. The molecules pass through a skimmer to the second chamber where we can use a time-of-flight mass spectrometer to analyze the molecular beam. The molecules traverse another skimmer to enter the third (science) chamber where the ion trap is located. The molecular beam and skimmers are aligned such that the molecules pass through the center of the ion trap. 

We use a linear segmented radiofrequency (rf) quadrupole trap \cite{willitsch12a} with a diagonal electrode distance of 3.5 mm to trap positively charged ions with trap frequencies ranging from 290 kHz ($m$=200 amu) to 1280 kHz ($m$=10 amu) along the longitudinal trap axis. The linear segmented design enables an effective harmonic trapping potential (see Fig. \ref{fig:exp}c) with an rf-free trapping region along its longitudinal axis \cite{Meir2017exp} tailored for precision spectroscopy of linear ion crystals \cite{Keller2016fix}. We optimized the trap for simultaneously trapping $^{40}$\Caion{} and $^{14}$\Nion{} in a two-ion linear crystal (see Fig. \ref{fig:IonsPics}e). We load \Caion{} into the trap from a skimmed atomic beam of neutral Ca atoms by a [1+1'] resonant ionization scheme using an external-cavity diode laser (ECDL) at 423 nm and a free-running diode laser at 375 nm \cite{lucas04a}.

The atomic-ion qubit used in the present scheme is comprised of the 4S$_{1/2}\left(m=-1/2\right)$ electronic ground state and the 3D$_{5/2}\left(m=-5/2\right)$ metastable state of \Caion{} (Fig. \ref{fig:N2lines}a) where $m$ denotes the magnetic quantum number. We drive the population between these states using an ECDL locked to an ultra-low expansion cavity at 729 nm. We detect the state of the atomic qubit by state-selective fluorescence using lasers at 397 nm and 866 nm driving the 4S$_{1/2}\leftrightarrow4$P$_{1/2}\leftrightarrow3$D$_{3/2}$ closed cycling transitions. We collect the fluorescence onto an electron-multiplying charge-coupled-device (EMCCD) camera and a single-photon counter using a high-numerical-aperture objective mounted in vacuum close to the ion trap (see Fig. \ref{fig:exp}b). 

\section{Preparation}\label{sec:prep}
\begin{figure}
	\centering
	\includegraphics[width=\linewidth,trim={1.4cm 6.5cm 7.6cm 1.5cm},clip]{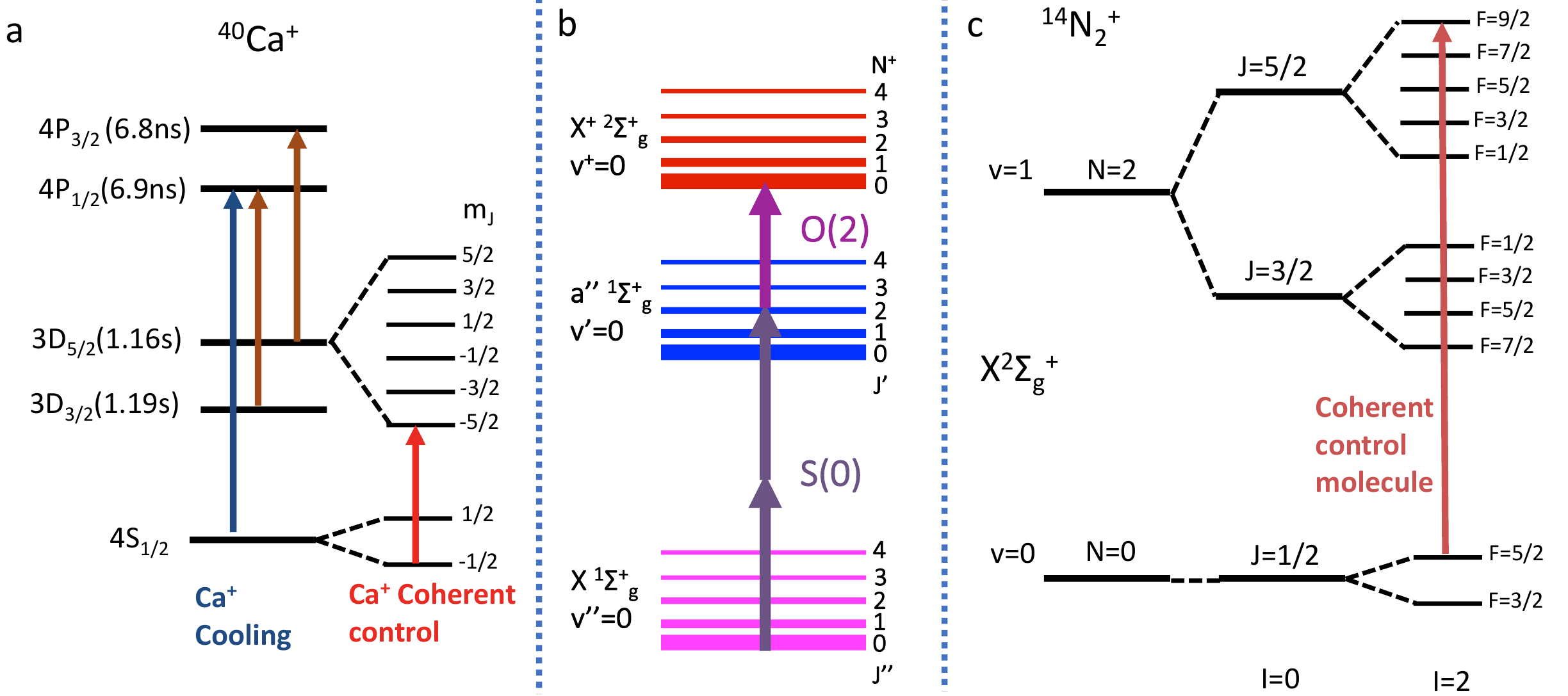}\\
	\caption{a) Energy levels of \Caion{} relevant for the present experiments (not to scale). Blue and red arrows indicate the transitions used for cooling and coherent manipulation, respectively. Brown arrows indicate repumping transitions. Numbers in brackets correspond to the radiative lifetimes of the relevant levels. b) [2+1'] REMPI scheme to create \Nion{ } ions in their rovibronic ground state \cite{tong10a}. Two photons at 202 nm (deep blue) couple the  $X~^1\Sigma_g^+~(v=0, N=0)$ level of N$_2$ to the intermediate $a''~^1\Sigma^+_g~(v=0,N=2)$ state. An additional photon at 375 nm (purple) ionizes the molecule to the $X~^2\Sigma_g^+~(v=0,N=0)$ ground state of \Nion{}. c) Energy levels of \Nion{} in its $X~^2\Sigma_g^+$ electronic ground state relevant for the present study (not to scale). The molecular quantum numbers used correspond to the molecular vibration ($v$=1,2,...), the electron spin ($S$=1/2), the molecular rotation ($N$=0,2,...), the spin-rotation structure ($\Vec{J}=\Vec{N}+\Vec{S}$) and the hyperfine structure ($\Vec{F}=\Vec{I}+\Vec{J}$). Here, we show only the ortho isomer with nuclear-spin quantum numbers $I$=0,2. Zeeman splittings due to an external magnetic field are not shown. The red arrow indicates a possible transition for coherent coupling using a quantum-cascade laser at 4.5 $\mu$m.}
	\label{fig:N2lines}
\end{figure}

We define a general pure molecular state by the following quantum numbers,
\begin{equation}\label{eq:Molstate}
    |\{v,N,S,J,I,F,m_F\}_i\rangle\equiv|\chi_i\rangle.
\end{equation}
Here, $v=0,1,...$ is the vibrational quantum number, $N=0,1,2,...$ the rotational quantum number, $S=1/2$ the electron-spin quantum number, $J=N+S, ..., N-S$ the quantum number of the total angular momentum without nuclear spin, $I=0,1,2$ the nuclear-spin quantum number, $F=J+I,..,J-I$ the quantum number of the total angular momentum and $m_F$ the associated magnetic quantum number of \Nion{ } (see Fig. \ref{fig:N2lines}c and Ref. \cite{germann16b}). We omit the electronic label (X $^2\Sigma^+_g$). The Greek letter $\chi$ denotes a set of quantum numbers where the index $i$ labels different states. 

We use a pulsed gas valve to create short, dense, rotationally, vibrationally and translationally (in the moving frame) cold packets of N$_2$ molecules. 
Resonance-enhanced multi-photon ionization (REMPI) is employed to ionize N$_2$ molecules from the molecular beam passing the center of the ion trap into a specific rovibrational cationic state following Refs. \cite{tong10a,tong11a}. A [2+1'] REMPI scheme employing ns pulsed laser beams at 202 and 375 nm to create \Nion{} molecules in the $|X~^2\Sigma_g^+;v=0;N=0\rangle$ rovibrational ground state is shown schematically in Fig. \ref{fig:N2lines}b.

Although the present REMPI scheme allows us to initialize the molecular ion in a specific rovibrational (quantum numbers $v$, $N$) state, the fine, hyperfine and Zeeman structure (quantum numbers $J$,$F$,$m_F$) are not resolved in the ionization step with the current laser system. Hyperfine-selective REMPI schemes can be devised using more narrow-bandwidth lasers \cite{GermannChemPhysI2016fix,GermannChemPhysII2016fix}. For the present purpose in order to prepare the molecule in a specific ($J, F, m_F$) state, a projective pumping scheme \cite{chou17a} is envisaged to be further discussed in Sec. \ref{sec:coherent}.

Mm-scale ion traps, such as the one used in our experiment, support large trapping volumes and trapping depths. For that, a molecular ion generated in the center of the trap will remain confined with almost unit probability. We synchronize a single ionization-laser pulse and a single molecule packet from the molecular beam to overlap in the trap center. We tune both the 202~nm and 375~nm laser intensities to achieve less than one ionization event on average per trial in order to reduce the probability of creating two ionization products in a single shot.


\section{Cooling}\label{sec:cooling}
Atoms with closed-cycling photon-scattering transitions such as the 4S$_{1/2}\leftrightarrow4$P$_{1/2}\leftrightarrow3$D$_{3/2}$ in \Caion{} shown in Fig. \ref{fig:N2lines}a are readily Doppler laser cooled to mK temperatures \cite{willitsch12a}. Using resolved-sideband cooling, they can further be cooled to the quantum ground state of their trapping potential. Molecules and atoms which lack these transitions cannot be cooled using direct laser cooling. In this context, sympathetic cooling (SC) has been proven to be an extremely efficient method allowing the preparation of cold samples of a variety of of atomic and molecular ions in traps \cite{molhave00a,willitsch12a}. 

In SC, auxiliary atomic ions are continuously laser cooled in the ion trap. Due to the long-range Coulomb interaction between the ions, energy is constantly exchanged between the hot molecular and the cold auxiliary ions in collisions. The continuous laser cooling of the auxiliary ions dissipates energy from the system until both species thermalize.

\begin{figure}
	\centering
	\includegraphics[width=\linewidth,trim={2.2cm 4.2cm 3.8cm 1cm},clip]{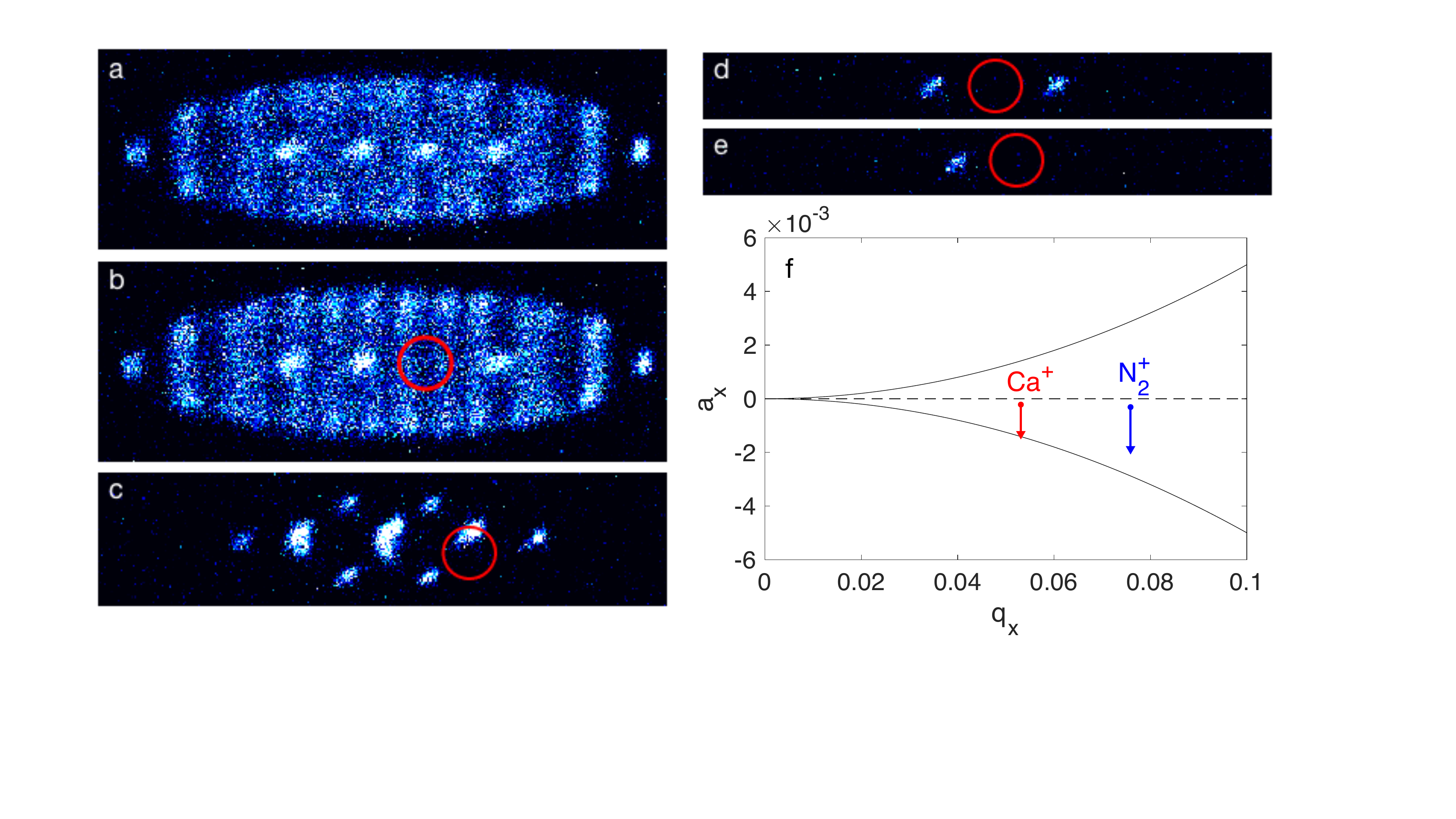}\\
	\caption{a) False-color fluorescence image of a pure crystal of $\sim$30 \Caion{} ions. b) The same crystal with an embedded additional single sympathetically-cooled non-fluorescing \Nion{} molecular ion. The red circle marks the position of the molecule. c-e). The same crystal after a sequence of evaporation steps to reduce the number of \Caion{} ions. The position of the molecule in the crystal is marked by a red circle. f) The position of \Nion{} (blue) and \Caion{} (red) in the stability diagram of the trap during the reduction sequence.}
	\label{fig:IonsPics}
\end{figure}

The SC rate is proportional to the number of auxiliary coolant ions \cite{Guggemos2015fix}. In our experiment, it takes a few minutes to cool a single \Nion{} ion with a single \Caion{} auxiliary coolant ion which is prohibitively long. Instead, we use a small crystal, consisting of roughly 30 \Caion{} ions (Fig. \ref{fig:IonsPics}a) to sympathetically cool a molecular ion within few seconds. 

Our method to read and coherently control the state of the molecule requires precise control over a single motional normal mode of the molecule-atom crystal. For that, it is preferable to work with a crystal made of a single molecule and a single atom. We reduce the number of auxiliary ions from about 30 to 1 (Figs. \ref{fig:IonsPics}b-e) by changing the trapping parameters for short periods of time such that the heavy \Caion{} ions are on the edge of the stability region of the ion trap \cite{major05a} while the light \Nion{} molecules are in a stable region (Fig. \ref{fig:IonsPics}f). By repeating this process while monitoring the number of \Caion{} ions in the trap, we can reduce the number of atomic ions quickly to a single one with high fidelity (Fig. \ref{fig:IonsPics}e). 

\begin{figure}
	\centering
	\includegraphics[width=\linewidth,trim={5.5cm 2cm 6.5cm 3.5cm},clip]{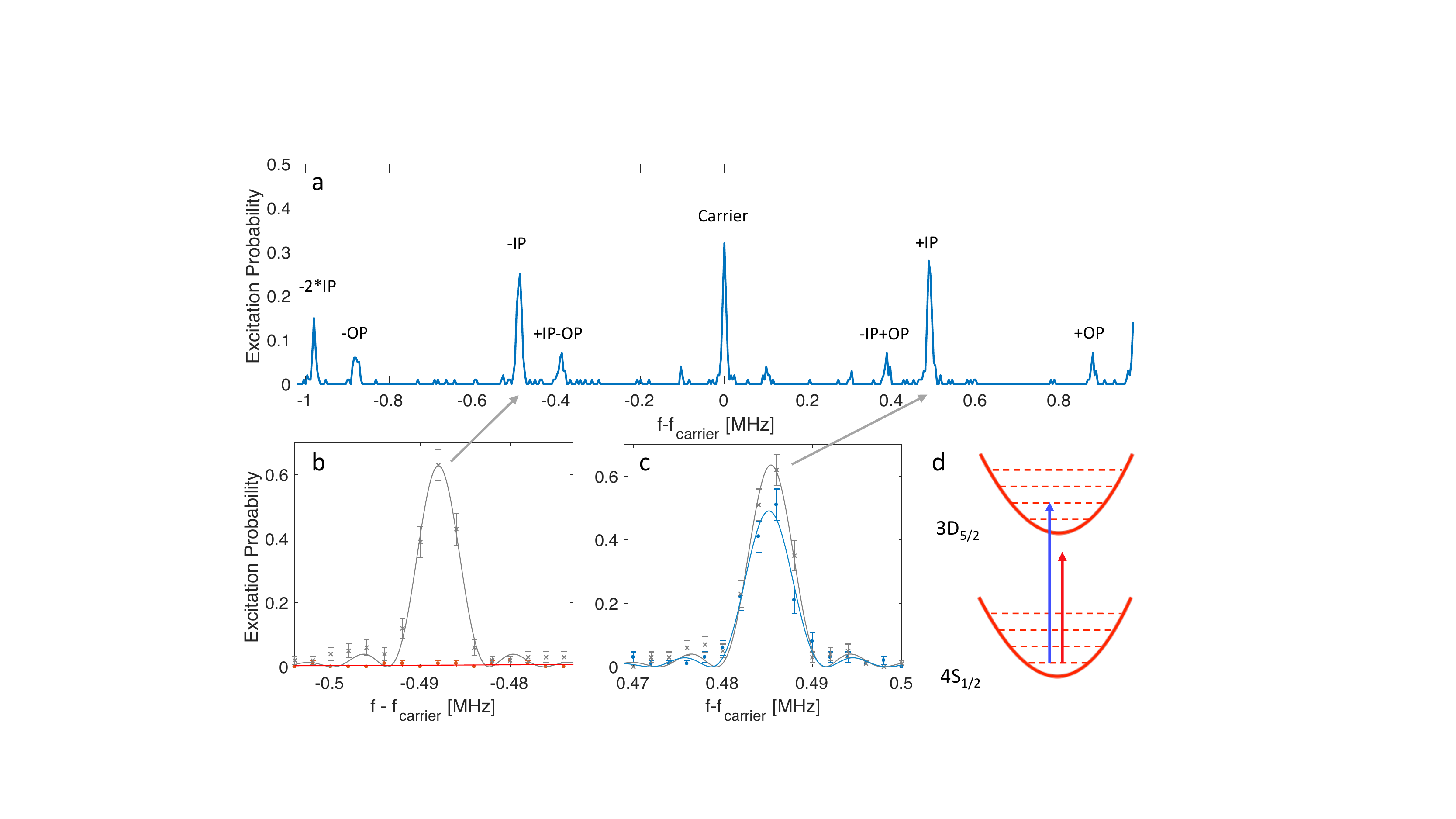}\\
	\caption{a) Excerpt of a resolved-sideband spectrum of a Doppler-cooled \Nion-\Caion{} two-ion crystal recorded on the 4S$_{1/2}\left(m=-1/2\right)\rightarrow$3D$_{5/2}\left(m=-5/2\right)$ quadrupole transition in \Caion{}. The spectrum shows excitations of motional modes of the crystal accompanying the electronic transition of the atomic ion. The origin of the frequency axis is the position of the "carrier" corresponding to the transition with $\Delta n=0$. The sidebands are designated by the relevant motional modes IP and OP, see text. b) and c) Magnified views of the first red (-IP) and blue (+IP) sidebands before (gray) and after (color) cooling to the motional ground state. Lines are fits to the data from which we extract a ground-state population of 98\%. d) Illustration of the red and blue sideband transitions originating from the motional ground state. A red sideband transition from this level (red arrow) can no longer be observed, as seen in the red trace in panel b.}
	\label{fig:GSCresults}
\end{figure}

In a subsequent stage, the two-ion string is cooled to the quantum ground state of one of its motional modes, typically the in-phase (IP) or out-of-phase (OP) mode \cite{morigi01a}, in the resolved-sideband regime. That is, using a narrow-linewidth laser at 729 nm the population is pumped to the motional ground state by addressing the individual levels of the quantum harmonic oscillator of the specific mode appearing as sidebands on an electronic transition of the atomic ion. A resolved-sideband spectrum of the narrow 4S$_{1/2}\left(m=-1/2\right)\rightarrow$3D$_{5/2}\left(m=-5/2\right)$ electric-quadrupole transition in \Caion{} (Fig. \ref{fig:N2lines}a) obtained after Doppler cooling is shown in Fig. \ref{fig:GSCresults}a. We use a series of second- and first-red-sideband pulses which change the motional quantum number by $\Delta n=-2$ and $\Delta n=-1$, respectively, to pump the IP mode of the ion-molecule crystal into its motional ground state. We reach 98\% ground-state population in a typical experiment (see Fig. \ref{fig:GSCresults}b-d).

\section{Detection}\label{sec:detection}
At this point, the shared motion of the atom-molecule crystal is in the ground state of the trapping potential denoted $|0\rangle$. The molecule is in a mixed state of different Zeeman, hyper-fine, spin-rotation and possibly even rotational levels depending on the precise characteristics of the REMPI scheme. The combined mixed molecular and pure motional states are then denoted as,
\begin{equation}
    \sum_{i}p_i|\chi_i\rangle\langle\chi_i|\otimes|0\rangle\langle0|.
\end{equation}
Here, $p_i$ is the classical probability that the molecule is in state $|\chi_i\rangle$ (Eq. \ref{eq:Molstate}) after the preparation step.

\begin{figure*}
	\centering
    \includegraphics[width=0.3\linewidth,trim={6cm 0cm 6cm 0cm},clip,angle=90]{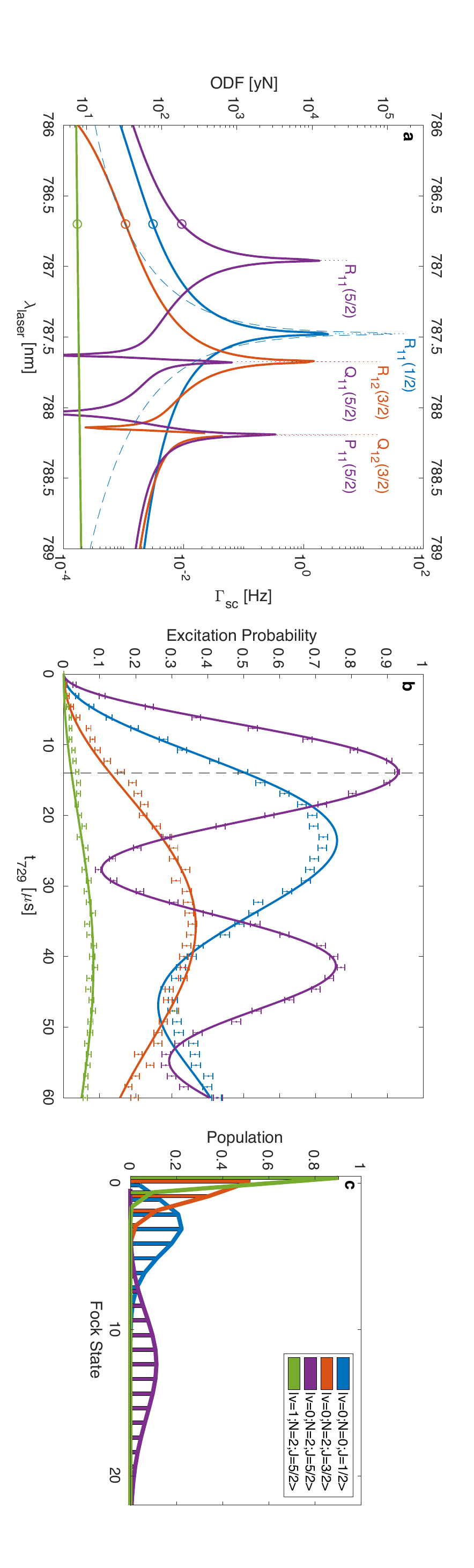}
	\caption{a) Theoretical calculations for the strength of the optical dipole force (ODF) used for coherent motional excitation as a function of the ODF-laser wavelength around the transition from the $X~^2\Sigma^+_g,v=0$ ground vibronic state to the $A^2\Pi^+_u,v=2$ excited vibronic state of \Nion{}. Solid lines are the amplitude of the ODF (in yN=$10^{-24}$N) for molecules in the states ($|v,N,J,m_J\rangle, I=0, S=1/2): |0,0,1/2,-1/2\rangle$ (blue), $|0,1,3/2,-1/2\rangle$, (red), $|0,2,5/2,-1/2\rangle$ (purple) and $|1,2,5/2,-1/2\rangle$ (green). The peaks correspond to rovibronic resonances in the X-A(2,0) band labeled using the spectroscopic notation of Ref. \cite{herzberg89a}. Circles mark a specific choice of the ODF-laser wavelength (786.5 nm) and the associated amplitudes of the ODF which are used in (b) for simulating the coherent-excitation signal on an \Nion{} ion. The dashed blue line represents the scattering rate for the $R_{11}$ transition originating from the molecular state $|0,0,1/2,-1/2\rangle$. A (single-beam) laser power of 15 mW was used in this simulation. b) Rabi flops of a single \Caion{} excited on the 4S$_{1/2}\left(m=-1/2\right)\rightarrow$3D$_{5/2}\left(m=-5/2\right)$ first-red-sideband transition following the application of an ODF which mimics the expected molecular signal according to the ODF amplitudes predicted in (a) and a pulse time of 0.75 ms. The color code used is the same as in (a) for different molecular states. Each point is the result of 400 experimental cycles. Error bars are 1$\sigma$ binomial errors. Lines are a fit to the function obtained for a red-sideband Rabi flop with a coherent distribution of Fock states \cite{leibfried03a,Meir2016} (see Appendix). The dashed line indicates a point of time in the population dynamics at which signals originating from different initial states are readily distinguishable. c) The coherent Fock-state distributions obtained from the fits to the data in (b) illustrating the different amplitudes of coherent motion generated by the ODF expected for \Nion{} in different rovibrational states.}
	\label{fig:N2Qsim}
\end{figure*}

To detect the state of the molecule, we use an optical-dipole force (ODF) to excite coherent motion, $|\alpha\rangle$, in the IP mode of the ion-molecule crystal. The ODF strength, and hence the amplitude of the coherent motion, depends on the molecular state through the frequency detuning of the ODF laser beam from a resonance in the molecule (see Fig. \ref{fig:N2Qsim} and the Appendix) \cite{foot05a} such that the motional state is now mixed with the molecular state,
\begin{equation}\label{eq:state}
    \sum_{i}p_i|\chi_i\rangle\langle\chi_i|\otimes|\alpha_i\rangle\langle\alpha_i|.    
\end{equation}
The problem of detecting the molecular state is now turned into detecting and differentiating between non-overlapping coherent motional states, $\langle\alpha_i|\alpha_j\rangle\neq1$.

To create the ODF, we use two counter-propagating Gaussian laser beams (10 mW, 25 $\mu$m waist) closely detuned from specific rotational transitions from the $X~^2\Sigma^+_g,v=0$ ground vibronic state to the $A^2\Pi^+_u,v=2$ excited vibronic state of \Nion{}. The beams create an intensity modulation pattern with a $\lambda/2$ period where $\lambda$=786.5 nm is the laser wavelength. We shift the frequency of the beams with respect to each other by the frequency of the IP mode of the ion-molecule crystal. This creates a time-dependent ac-Stark-shift modulation that acts as a resonant driving force exciting the IP mode \cite{Meekhof1996}. 

In Fig. \ref{fig:N2Qsim}a, we show the calculated amplitudes of the ODF for \Nion{} ions in different spin-rotational states of the vibrational ground and first excited state as function of the ODF-laser wavelength (see the Appendix for details of the calculation). The traces are labeled by the quantum number $J$ given in brackets. The ODF shows marked enhancements around resonance for rovibronic transitions labeled using the usual spectroscopic notation \cite{herzberg89a}. The ODF amplitude decreases as $\sim 1/\Delta$, where $\Delta$ is the detuning of the laser frequency from a resonance. A major loss channel of the molecular state is off-resonant excitation on the transition used to enhance the strength of the ODF at detuning $\Delta$. The scattering rate \cite{foot05a} scales as $\sim 1/\Delta^2$, see the blue dashed line in Fig. \ref{fig:N2Qsim}a for the example of the $R_{11}(1/2)$ transition. Thus, in a practical situation the frequency of the ODF laser beams should be tuned close enough to a specific transition to ensure a strong ODF, but remain detuned far enough so as to decrease the probability for off-resonant scattering.

The ODF excites coherent motion of the \Caion{}-\Nion{} crystal that depends on its amplitude and pulse time. To read the coherent motional state of the two-ion crystal, Rabi sideband thermometry is employed \cite{leibfried03a}. That is, Rabi oscillations are excited on a red sideband of the \Caion{} 4S$_{1/2}\rightarrow$3D$_{5/2}$ transition using the narrow 729~nm laser. The frequency and contrast of the Rabi oscillations depend on the populations of the Fock, i.e., harmonic oscillator, states populated by coherent excitation from the ODF. If no motional excitation has occurred, no red-sideband transitions are possible from the motional ground state (see Fig. \ref{fig:GSCresults}c) and, therefore, no Rabi flops can be observed.

In Fig. \ref{fig:N2Qsim}b, we used a single \Caion{} ion to experimentally simulate the signal expected for \Nion{} in different rovibrational states shown in Fig. \ref{fig:N2Qsim}a (see Appendix for details of the simulation). Such "quantum simulations" of the molecular signal using a purely atomic system are useful to characterize and optimize the methodology for state readout. 
The results are shown in Fig. \ref{fig:N2Qsim}b which shows Rabi flops on the 4S$_{1/2}\left(m=-1/2\right)\rightarrow$3D$_{5/2}\left(m=-5/2\right)$ first-red-sideband transition in \Caion{} obtained after 0.75 ms action of the ODF. The different traces show experiments in which the ODF laser power was set to mimic the signal expected for \Nion{} in the I=0, S=1/2, $|v,N,J,m_J\rangle$= $|0,0,1/2,-1/2\rangle$, $|0,1,3/2,-1/2\rangle$, $|0,2,5/2,-1/2\rangle$ and $|1,2,5/2,-1/2\rangle$ rovibrational states of the electronic ground state. The period and contrast of the signal depend on the amplitude of the coherent motional state generated by the action of the ODF which can be reconstructed from these signals \cite{leibfried03a,Meir2016}. The underlying distribution of Fock states extracted from a fit to the data (solid lines in Fig. \ref{fig:N2Qsim}b) is shown in Fig. \ref{fig:N2Qsim}c. We see that even for small differences in the amplitude of the ODF and the overlapping coherent motional distributions, the Rabi flops are distinguishable. This clearly illustrates that the method can be used to discriminate different rotational states of the molecular ion.     

\section{Coherent manipulation}\label{sec:coherent}
After describing our state detection method, we briefly consider the coherent coupling of two molecular states using a narrow-bandwidth laser, for instance specific hyperfine-Zeeman levels of rotational components of the first two vibrational states such as $\left|\chi_i(v=0,N=0)\right\rangle\rightarrow\left|\chi_j(v=1,N=2)\right\rangle$ shown in Fig. \ref{fig:N2lines}c.
The transitions are around 4.5 $\mu$m (65 THz) and are dipole forbidden such that they are weak and narrow \cite{germann14a}. These transitions are of interest due to their unique properties for clock applications and spectroscopic tests of a possible temporal variation of the ratio of the proton and electron mass \cite{kajita14a,kajita15a}.

\begin{figure} 
	\centering
	\includegraphics[width=\linewidth,trim={1cm 0cm 7cm 5cm},clip]{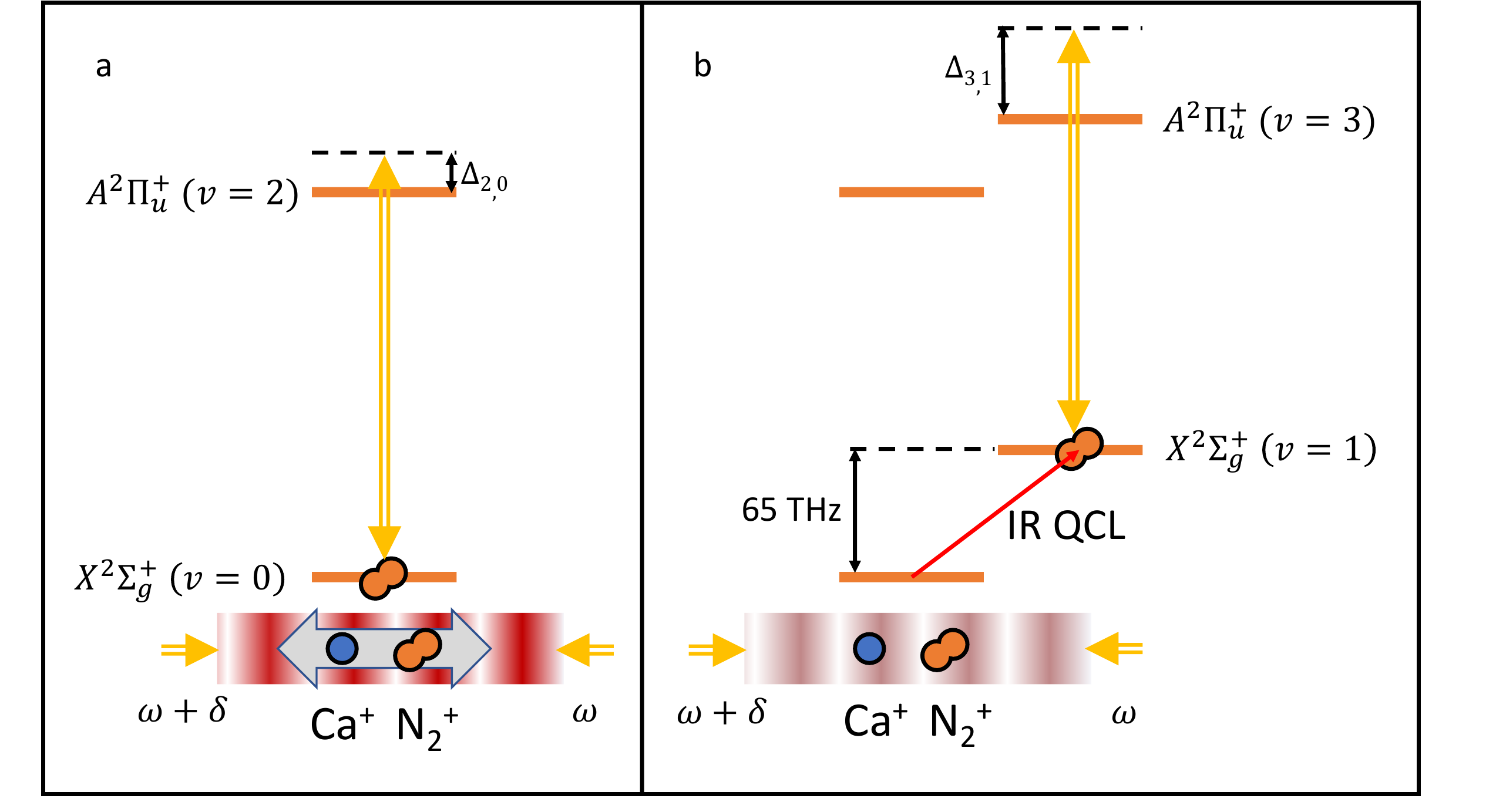}\\
	\caption{Simplified energy-level scheme of \Nion{} to illustrate the effect of vibrational excitation on the ODF. a) \Nion{} is in the vibronic ground state and the ODF is detuned by $\Delta_{2,0}$ from the (2,0) band of the \AX{} transition. b) When the molecule is excited to the v=1 excited vibrational state, the detuning of the ODF from the nearest vibronic transition (in this case the (3,1) band of the \AX{} transition) increases significantly resulting in a substantial reduction of the ODF signal.}
	\label{fig:N2_ODF}
\end{figure}

A coherent pulse on this infrared (IR) transition creates a superposition of the vibrational ground and first excited states,
\begin{equation}
    \left(a\left|\chi_i\right\rangle + b\left|\chi_j\right\rangle\right) \otimes 
    \left|0\right\rangle.
\end{equation}
As described in Sec. \ref{sec:detection}, an ODF pulse can be applied that entangles coherent motion, $\left|\alpha\right\rangle$, with the molecular state. The motional amplitude is much bigger for the vibrational ground state than for the first vibrationally excited state due to the increase in detuning of 7.2 THz of the (3,1) compared to the (2,0) vibronic band in the \AX{} system (Fig. \ref{fig:N2_ODF} and Fig. \ref{fig:N2Qsim} purple and green traces). The resulting molecular and motional state is:
\begin{equation}
    a\left|\chi_i\right\rangle \otimes 
    \left|\alpha_i\right\rangle + b\left|\chi_j\right\rangle \otimes 
    \left|\alpha_j\right\rangle.
\end{equation}
Upon detection of the coherent state $|\alpha_i\rangle$ or $|\alpha_j\rangle$, the molecule collapses either into the $\left|\chi_i\right\rangle$ or the $\left|\chi_j\right\rangle$ state with probability $|a|^2$ and $|b|^2$, respectively. The signal may not be strong enough to detect the molecular state in a single experimental cycle. However, since the molecule collapsed into a specific state upon measurement of the coherent state, we can recool the motion and repeat the detection cycle as many times as needed for a high-fidelity state readout \cite{Hume2007fix}. By recording $|b|^2$ as a function of the spectroscopy-laser frequency, a spectrum of the IR transition can be obtained.

The coherent coupling between two different vibrational states can also be used to project the molecule into a specific spin-rotation, hyperfine and Zeeman level. This is useful, for instance, if this specific state cannot be uniquely prepared in the REMPI step because of a lack of resolution of the ionization lasers. Consider a narrow-bandwidth IR laser which couples two levels of the ground and excited vibrational states, $\left|\chi_i\right\rangle \rightarrow \left|\chi_j\right\rangle$. Now assume that the state $\left|\chi_i\right\rangle$ is initially not occupied such that the IR laser does not affect the molecule. Additional Raman, rf and microwave coupling can be used to mix different states in $v=0$, including any populated level with $\left|\chi_i\right\rangle$. This will then result in a coupling to the $\left|\chi_j\right\rangle$ level by the IR laser and ODF detection will project the molecule into one of the coupled states. Thus, while it may not be possible to reliably disentangle different spin-rotation-hyperfine-Zeeman levels in the ground vibrational state by the ODF, a projection into the $v=1$ state can readily be identified with high fidelity. Therefore, if coherent-excitation signal associated with $v=1$ is detected, then the molecule is necessarily prepared in the specific state $\left|\chi_j\right\rangle$ addressed by the IR laser. 


\section{Experimental errors}\label{sec:error}
Several sources of error may compromise the fidelity of the present scheme. The most important ones will be discussed in the following section.

\subsection{Chemical lifetime of \Nion}
Chemical reactions of the single molecular ion with background gas molecules such as H$_2$ represent a major problem even under ultrahigh-vacuum conditions. In our experiment with an H$_2$ background pressure of $1.7\cdot10^{-10}$ mbar, we measured the $(1/e)$ chemical life time of \Nion{} to be about 5 minutes until \Nion{} reacts with H$_{2}$ to form N$_2$H$^{+}$. We detect this chemical product using mass spectrometry in the ion trap \cite{drewsen04a}. Due to this relatively short chemical lifetime of \Nion{}, it is important to keep the time from ionization to the beginning of the actual experiments as short as possible. We found that the reduction procedure illustrated in Fig. \ref{fig:IonsPics} and Sec. \ref{sec:cooling} can produce \Caion{}-\Nion{} two-ion crystals reproducibly within 1-2 min. This leaves a few minutes for the actual experiments. Since every experimental cycle takes about 20 ms, this allows tens of thousands of measurements on the same molecule within its chemical lifetime, where typically few 100s should be enough to detect the molecular state with high fidelity (see the dashed black line in Fig. \ref{fig:N2Qsim}b). 

\subsection{Radiative lifetime of \Nion}
Off-resonant photon scattering from the ODF beams, i.e., excitation and subsequent fluorescence on some component of the \AX{} transition of \Nion, will lead to the loss of the initially prepared quantum state of \Nion{} and to the possible population of highly excited vibrational levels of the electronic ground state which do not couple to the ODF. This terminates the experimental sequence and necessitates the re-initialization of the experiment. The scattering rate depends on the detuning of the ODF laser beams (see dashed blue line in Fig. \ref{fig:N2Qsim}a). Smaller detunings will allow for a better distinguishability between different molecular states due to an enhanced ODF strength, however, at the expense of a shorter radiative lifetime. For the rotational ground state, and our chosen laser wavelength, the theoretical scattering rate is 0.5 mHz. This means that for an ODF pulse time of 0.75 ms used in our experiment, we can expect to perform more than 2 million experimental cycles before scattering a single photon (see Appendix for details on the scattering-rate calculations).

\subsection{Background signal from coherent motional excitation on \Caion}
\begin{figure}
	\centering
	\includegraphics[width=\linewidth,trim={3.2cm 2cm 3.6cm 2cm},clip]{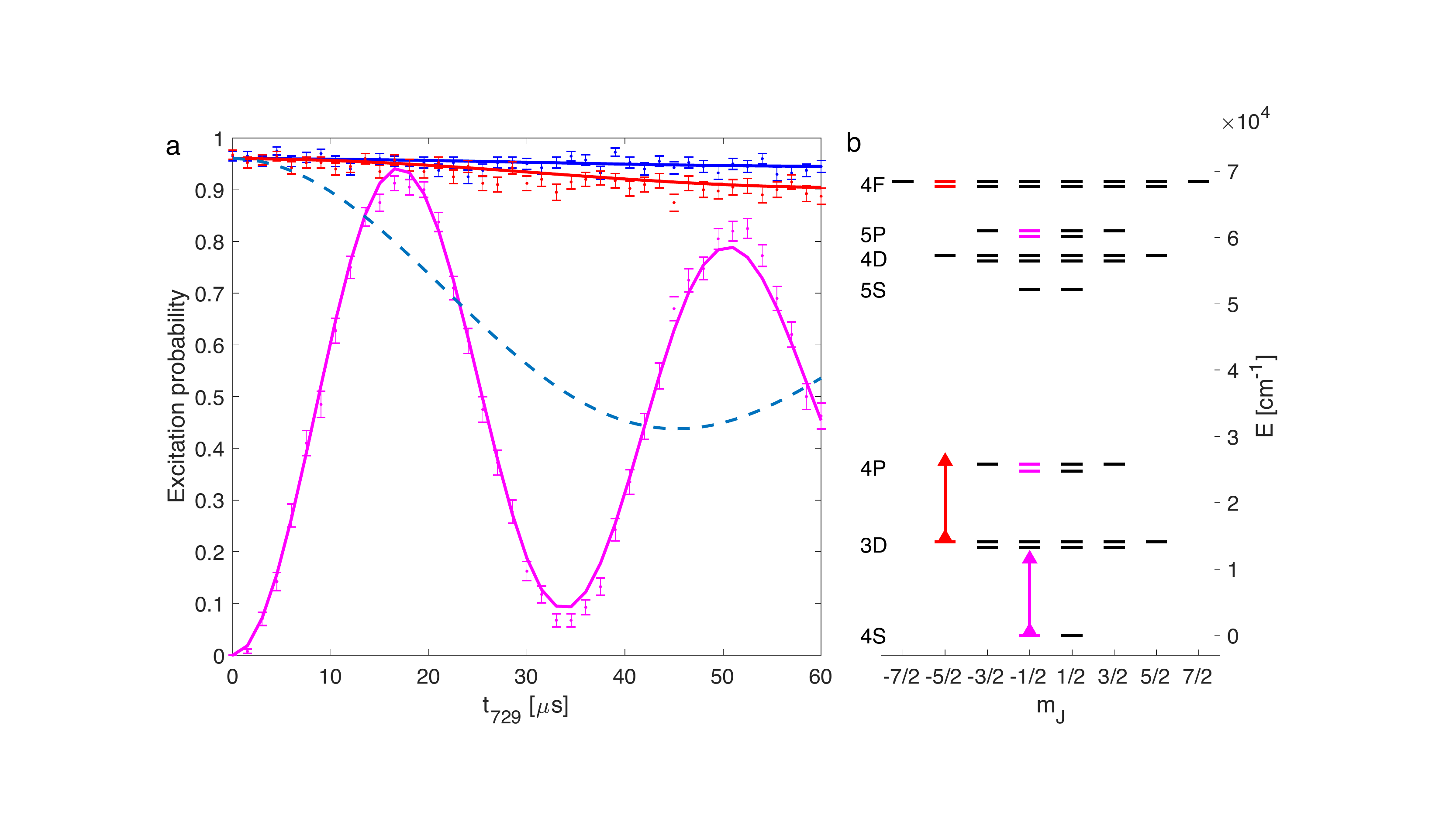}\\
	\caption{a) Rabi flops of a single \Caion{} ion excited on the 4S$_{1/2}\left(m=-1/2\right)\rightarrow$3D$_{5/2}\left(m=-5/2\right)$ transition. Magenta trace: the ion was prepared in the 4S$_{1/2}\left(m=-1/2\right)$ state and an ODF pulse of 0.75 ms was applied prior to the red-sideband spectroscopy (same data as shown in the purple trace of Fig. \ref{fig:N2Qsim}b). Red trace: blue-sideband Rabi signal after the same ODF pulse but with the ion initialized in the 3D$_{5/2}\left(m=-5/2\right)$ state. Blue trace: same as the red trace, but without any force applied to the ion. The ODF single-beam laser power employed was 10 mW. Dashed line: expected signal for an \Nion{} ion in the state $|v=0,N=0,J=1/2,m_J=-1/2\rangle$ where the \Caion{} is initialized in the 3D$_{5/2}\left(m=-5/2\right)$ level to suppress background signal. b) Energy levels including Zeeman sub-levels and spin-orbit levels (not to scale) of \Caion{}. Arrows indicate the ODF laser which couples levels of the same color (magenta: starting in the 4S$_{1/2}\left(m=-1/2\right)$ state, red: starting in the 3D$_{5/2}\left(m=-5/2\right)$ state) for linear polarization parallel to the external magnetic field.}
	\label{fig:HideCa}
\end{figure}

For the \Caion{} ion, the ODF laser beams are detuned by 375 THz from the strong $4S_{1/2}\rightarrow 4P_{1/2}$ cycling transition (Fig. \ref{fig:N2lines}a). Even so, the coupling between the S and P states is strong enough at this detuning to efficiently excite coherent motion by the ODF. We used this feature to simulate the signal of a molecule using \Caion{} in Fig. \ref{fig:N2Qsim}. However, in a real experiment this excitation serves as unwanted background noise for the molecular state detection since it is not dependent on the molecular state. To reduce this noise, we prepare the \Caion{} ion in the D$_{5/2}\left(m=-5/2\right)$ state and tune the polarization of the ODF beams parallel to the magnetic field quantization axis to decouple the ODF from the \Caion{} electronic states (see Fig. \ref{fig:HideCa}b). In this state, the residual ODF amplitude on \Caion{} is determined by coupling to highly excited $F$ states and polarization imperfections. We measured its strength to be much lower than the expected molecular signal (see Fig. \ref{fig:HideCa}a).

\section{Outlook}\label{sec:outlook}
In this manuscript, we describe an experimental approach for preparing, detecting and coherently manipulating a single molecular ion. We demonstrated the preparation of a \Caion{}-\Nion{} two-ion crystal in the ground-state of their combined motion in an ion trap, on demand and in a short time compared to the chemical life time of the molecule. We also demonstrated a new experimental technique to non-destructively detect the state of the molecule. The method can be conveniently optimized using a single atomic ion to simulate the molecular signal strength and illustrate its sensitivity to both the vibrational and rotational state of the molecule at a specific detuning of the ODF laser beam. We also discussed experimental aspects of trapping and manipulating single molecules, such as the sympathetic cooling time, the chemical life time, the radiative life time and factors impacting the signal-to-noise-level of our detection method. 

While all the foundations for the present QLS detection of single molecules were characterized in this manuscript, the observation and analysis of this signal with \Nion{} ions is currently carried out and is outside the scope of this manuscript.

The prospects of the present quantum-logic manipulation of single molecules are overwhelming. The ability to perform precision spectroscopy on a single molecule prepared in a specific quantum state will open up opportunities for precise tests of molecular theory \cite{mur-petit12a}, for new frequency standards in the mid-IR domain \cite{karr14afix,Schiller2014fix} and for testing fundamental theories which transcend beyond the standard model \cite{salumbides13a}. These types of experiments can also open up new avenues in chemical studies in which, e.g., collisions and reactions of a single molecule in a specific quantum state are monitored in real time.

While our work currently focuses on \Nion{}, the methods shown here are not limited to a specific type of molecular ion, but can be adopted to a wide class of diatomic and polyatomic species.

\section*{Acknowledgements}

We acknowledge funding from the Swiss National Science Foundation as part of the National Centre of Competence in Research, Quantum Science and Technology (NCCR-QSIT) and the University of Basel. 



\bibliographystyle{apsrev4-1}

\newpage

\onecolumngrid

\section*{Appendix}

\renewcommand{\thesubsection}{A\arabic{subsection}}

In the following, we provide further details on the calculations of the ac Stark shifts and the derived optical-dipole forces (ODF) acting on the ions, the modelling of the coherent motional excitation of the ions, the detection of coherent motional excitation by sideband thermometry, the simulation of the state-dependent coherent motional excitation of \Nion{} using \Caion and the scattering rate for parasitic off-resonant excitation of \Nion{}.

\subsection{Ac Stark shift of the molecular state due to the optical-dipole-force laser beams}
\label{app:acstark}

The characterization of the effect of the ODF on both \Caion{} and \Nion{} requires the calculations of the relevant ac Stark shifts. The ac Stark shift of a state $\left|i\right\rangle$ due to electric dipole transitions (E1) to all accessible states $\left|j\right\rangle$ in the case of large detunings, \textit{i.e.} $\Delta_j ^{(i)} \gg \Omega_j^{(i)},\Gamma^{(j)}$, is given by\cite{Grimm1999}:
\begin{equation}\label{eq:acstark}
    \Delta E_{ac}^{(i)} = \frac{\hbar}{4} \sum_j \sigma_j^{(i)} \frac{\left(\Omega_j^{(i)}\right)^2}{\Delta_j^{(i)}} = \frac{1}{2\epsilon_0 \hbar c}I\left(\textbf{x},t\right) \sum_j \sigma_j^{(i)}\frac{\left|\left\langle j \left| \bm{\mu} \right| i\right\rangle \right|^2}{\Delta_{j}^{(i)}}.
\end{equation} 
Here, $\Delta E_{ac}^{(i)}$ is the ac-stark shift in Joules, $\Omega_j^{(i)}$ the Rabi frequency, $I(\bm{x},t)$ the laser intensity, $\sigma_j^{(i)} = \pm 1$ for $E^{(i)}\lessgtr E^{(j)}$, $\left| \left\langle j \left| \bm{\mu} \right| i\right\rangle\right|$ the transition dipole moment matrix element and $\Delta_{j}^{(i)} = \omega_{laser} - \omega_{j}^{(i)}$ the detuning of the laser frequency relative to the transition frequency.

In the case of an external magnetic field $\bm{B} = \left(0,0,B_0\right)$ which removes the degeneracy of the $m_J$ states and for given rovibronic states $\{ \left| i\right\rangle, \left| j\right\rangle\} = \left| \psi_{el,vib, rot},J,m_J\right\rangle$, the transition dipole moment can be explicitly written as\cite{James1998, MatthiasThesis}:
\begin{equation}\begin{split}
    \left| \left\langle j \left| \bm{\mu} \right| i\right\rangle\right|^2 =\left|\left\langle j \left| \mu_{red} \right| i\right\rangle \right|^2
   \left| \sum_{q=-1}^{1}\begin{pmatrix}
J^{(i)} & 1 & J^{(j)}\\ 
-m_J^{(i)}& q & m_J^{(j)}
\end{pmatrix} \textbf{c}^{(q)}\bullet \bm{\varepsilon}\right|^2.
\end{split}\end{equation}
Here, $\left|\left\langle j \left| \mu_{red} \right| i\right\rangle \right|$ is the reduced matrix element, the term in brackets denotes a Wigner 3j-symbol, $\bm{\varepsilon}$ is the complex laser polarization vector and $\textbf{c}^{(q)}$ are normalized spherical basis vectors\cite{James1998}:
\begin{align}
    \textbf{c}^{(+1)}&=-\tfrac{1}{\sqrt{2}} (1,-i,0),\\
    \textbf{c}^{(0)}&=(0,0,1),\\
    \textbf{c}^{(-1)}&=\tfrac{1}{\sqrt{2}} (1,i,0).
\end{align}
In our experimental configuration, where the laser polarization is parallel to the external magnetic field, $\textbf{c}^{(q)}\bullet \bm{\varepsilon}=\delta_{q0}$ where $\delta_{ij}$ is the Kronecker delta.

The reduced matrix element can be calculated from experimentally measurable quantities by\cite{BernathSpectraAtomsMolecules}:
\begin{equation}
    \left|\left\langle j \left| \mu_{red} \right| i\right\rangle \right|^2 = \frac{3\epsilon_0 h c^3}{2 \left(\omega_j^{(i)}\right)^3}\left(2J^{(j)}+1\right) S_{J^{(i)}}^{(\Delta J)} A_{vib}^{(i,j)}.
\end{equation}
In here, $S_{J^{(i)}}^{(\Delta J)}$ is the H\"onl-London factor and $A_{vib}^{(i,j)}$ is the vibronic Einstein A-coefficient.

In the case of \Nion{} and a laser wavelength of $\lambda = 786.7\ \textrm{nm}$ close to the $(2,0)$-band of the first electronic transition \AX\cite{Wu2007}, we can neglect contributions of other transitions and restrict our calculations to this particular vibronic transition. This is justified by the fact that the \AX electronic excitation is comparably weak with vibronic Einstein A-coefficients on the order of $50\ \textrm{kHz}$\cite{Langhoff1987} and thus other vibrational bands, separated by $\sim66\ \textrm{THz}$\cite{germann14a} give negligible contributions. Further, the second electronic band $B^2\Sigma_u^+ \leftarrow X^2\Sigma_g^+$ around $391\ \textrm{nm}$\cite{Lofthus1977}, though stronger with an Einstein A-coefficient of $A_{vib}^{(0,0)} = 9.64 \ \textrm{MHz}$\cite{Lofthus1977}, is too far detuned to give a significant contribution. Moreover, its contribution is independent of the occupied rovibrational level due to the large detuning.

The $A^2\Pi_u$ excited state can best be described by a Hund's case (a) coupling scheme.  The $X^2\Sigma_g^+$ ground state is adequately described by a Hund's case (b) coupling scheme. This implies multiple allowed transitions from every rovibronic state of interest. For each transition, the reduced matrix elements were calculated neglecting nuclear spin, \textit{i.e.} $I=0$, using the $A_{vib}^{(i,j)}$ value from Ref.\cite{Langhoff1987}. The $S_{J^{(i)}}^{(\Delta J)}$ were calculated following Ref.\cite{Earls1935} with appropriate normalization. Spectroscopic constants were taken from Ref.\cite{Miller1984} and transition frequencies from Ref.\cite{Wu2007}. 

For the $R_{11}(1/2)$, $m_{J"} = m_{J'} = -1/2$ $\pi$-transition from the rovibrational ground state of \Nion{} and a laser beam with $I = 2P/ \pi w_0^2 = 10\cdot10^{6}\  \textrm{W}/\textrm{m}^2$ at $786.7\ \textrm{nm}$, we obtain an ac-stark shift of $\Delta E_{ac} \approx 0.9\ 10^{-30}\ \textrm{J} = h\cdot 1.3\ \textrm{kHz}$. 

\subsection{Calculation of the optical-dipole force acting on N$_2^+$ in a specific quantum state}
\label{app:odf}

The ODF is created from the interference of two counter-propagating laser beams with a Gaussian profile. The beams' waist ($w_0\sim$25 $\mu$m) is aligned to the center of the trap. The beams' Rayleigh range, $z_R=\pi w_0^2/\lambda\approx2.5$ mm with $\lambda=786.5$ nm is much larger than the distance of the ions from center of the trap which is few $\mu$m for a two-ion crystal. Thus, we can approximate the laser beams as plane waves in the region of the ions. 

A plane-wave, $\textbf{E}_{+}=\hat{e}E_0e^{+ikx-i\omega t+i\phi_{+}}$, is interfered with a counter propagating plane-wave, $\textbf{E}_{-}=\hat{e}E_0e^{-ik'x-i\omega' t+i\phi_{-}}$, of the same amplitude and polarization to form an intensity modulation:
\begin{equation}
    I(x,t)=\left|\textbf{E}_{+}+\textbf{E}_{-}\right|^2/2\eta\approx2I_0\left(1+\cos\left(2kx-\Delta\omega t+\Delta\phi  \right)\right).
\end{equation}
Here, $I_0=\left|\textbf{E}_\pm\right|^2/2\eta=2P/\pi w_0^2$ is the peak intensity of a single Gaussian beam where $\eta$ is the vacuum wave impedance, $P=10$ mW is the single-beam laser power, $k=2\pi/\lambda=\omega/c$ is the laser k-vector, $\omega/2\pi$ is the laser frequency, $\phi_\pm$ is the laser phase and $x$ is the position relative to the center of the trap. We tune the frequency difference between the two laser beams, $\Delta\omega=\omega-\omega'$, to match the frequency of the mode of a single ion or two-ion crystal in the trap. The phase difference of the laser beams is $\Delta\phi=\phi_+-\phi_-$ and we approximated $k+k'\approx2k$ since $\Delta\omega\ll\omega$.

The ODF is then given by:
\begin{equation}\label{eq:ODF}
    F_{ODF}(x,t)=-\dv{\Delta E_{ac}^{(i)}}{x}=-4 k \Delta E_{ac,0}^{(i)} \sin\left(2kx-\Delta\omega t+\Delta\phi\right).
\end{equation}
Here, $\Delta E_{ac,0}^{(i)}$ is the ac-Stark shift Eq. (\ref{eq:acstark}) calculated for a single lattice beam with intensity $I_0$.

\subsection{Numerical calculation of the coherent motional excitation of a 
single ion and a two-ion crystal by the optical-dipole force}
\label{app:cme}

After establishing the time-modulated potential and the resulting ODF acting on the ion, we turn to calculate the motional excitation dynamics due to this force. We treat only the single mode which the laser modulation is in resonance with. The classical equation of motion for the single-ion case is given by,
\begin{equation}\label{eq:eomsingle}
    m\ddot{x}=-k_t x +F_{ODF}(x,t).
\end{equation}
Here, $m$ is the mass of the ion, $x$ its position relative to the center of the trap and $k_t=m\omega_t^2$ characterizes the trap harmonic confinement which is determined experimentally by measuring the single ion harmonic frequency, $\omega_t/2\pi=641.15(5)$ kHz. We numerically integrate Eq. (\ref{eq:eomsingle}) using a 4$^\textrm{th}$ order Runge-Kutta method with initial conditions $x(0)=\dot{x}(0)=0$. The total energy of the ion is given by,
\begin{equation}\label{eq:Eone}
    E_{1}(t)=\frac{1}{2}\left(k_t x(t)^2+m\dot{x}(t)^2\right).
\end{equation}

For the two-ion case, the ions also repel each other by the Coulomb force so that
\begin{equation}\label{eq:eomtwo}
\begin{split}
    &m_1\ddot{x}_1=-k_t x_1 + \textrm{Sign}\left(x_1-x_2\right)\cdot \frac{e^2/4\pi\varepsilon_0}{\left|x_1-x_2\right|^2} + F_{ODF}(\Delta E_1,x_1,t)\\
    &m_2\ddot{x}_2=-k_t x_2 + \textrm{Sign}\left(x_2-x_1\right)\cdot \frac{e^2/4\pi\varepsilon_0}{\left|x_2-x_1\right|^2} + F_{ODF}(\Delta E_2,x_2,t).
\end{split}
\end{equation}
Here, each ion is labeled by a different index ($1,2$), $e$ is the elementary charge (we assume both ions are singly positively charged) and $\varepsilon_0$ is the vacuum permittivity. We allow for different strengths and directions of the ODF on each ion charectrized by the individual ac-Stark shifts $\Delta E_1$ and $\Delta E_2$.

The solution with the minimum energy of Eq. (\ref{eq:eomtwo}) with the ODF strength set to zero gives the equilibrium positions of the two ions \cite{morigi01a},
\begin{equation}
    x_0=\pm\left(\frac{1}{4}\frac{e^2/4\pi\varepsilon_0}{k_t}\right)^{1/3}.
\end{equation}
The Fourier transform of Eq. (\ref{eq:eomtwo}) (again with the ODF strength set to zero) gives the frequencies of the in-phase (IP) and out-of-phase (OP) normal modes which can also be calculated analytically\cite{morigi01a},
\begin{equation}
    \omega_\pm^2=\frac{k_t}{m_1}\left(1+1/\mu\pm\sqrt{1+1/\mu^2-1/\mu}\right).
\end{equation}
Here, $\omega_+$ is the OP and $\omega_-$ is the IP mode and $\mu=m_2/m_1\ge1$. For a Ca$^+$-N$_2^+$ two-ion crystal in our experiment, the IP mode frequency is $\omega_-/2\pi=690.2$ kHz.

The total energy of the two-ion system is given by,
\begin{equation}\label{eq:Etwo}
    E_{2}(t)=\frac{1}{2}\sum_{i=1,2}\left(k_t x_i(t)^2+m\dot{x}_i(t)^2\right) +
    \frac{e^2/4\pi\varepsilon_0}{\left|x_1(t)-x_2(t)\right|}-E_0.
\end{equation}
It is comprised of the kinetic and potential energies of the two ions in the harmonic trap and their mutual Coulomb energy. The minimal energy of the system is normalized to zero by the last term: $E_0=2\cdot\frac{1}{2}k_tx_0^2+\frac{e^2/4\pi\varepsilon_0}{2\left|x_0\right|}=3k_tx_0^2$.

\subsection{Side-band thermometry of single and two-ion crystal to extract the amplitude of the coherent motion}
\label{app:sbt}

A classical harmonic mode with energy $E$ is assumed to correspond to a coherent (Poisson) distribution of populations of Fock states:
\begin{equation}
    P(n;\bar{n})=e^{-\bar{n}}\frac{\bar{n}^n}{n!}.
\end{equation}
The mean phonon number of the distribution, $\bar{n}$, is determined from the classical energy by $E=\bar{n}\hbar\omega_*$. Here, $\omega_*$ is either the single-ion motional frequency, $\omega_t$, for the single-ion case or the IP mode frequency, $\omega_-$, for the two-ion case.

The red-sideband Rabi signal on the 4S$_{1/2}\left(m=-1/2\right)\rightarrow$3D$_{5/2}\left(m=-5/2\right)$ transition in Ca$^+$ is given by a sum over the different Rabi oscillations for the different Fock states weighted by their populations \cite{leibfried03a},
\begin{equation}\label{eq:Rabi}
    P_e(t)=\sum_n P(n;\bar{n}) \frac{\Omega_{n,n-1}^2}{\Omega_{n,n-1}^2+\Delta^2}\sin^2\left(\frac{1}{2}\sqrt{\Omega_{n,n-1}^2+\Delta^2} \cdot t\right).    
\end{equation}
Here, $\Delta$ is the laser detuning from the sideband transition. The Rabi frequency for a red sideband transition from harmonic oscillator level $n$ to $n-1$ is given by, $\Omega_{n,n-1}=\Omega_0e^{-\eta^2/2}\eta L_{n-1}^1(\eta^2)/\sqrt{n}$, where $\Omega_0\approx\frac{\pi}{7~\mu\textrm{s}}$ is the bare Rabi frequency which we determine experimentally, $L_n^1(x)$ is a generalized Laguerre polynomial and $\eta$ is the Lamb-Dicke parameter\cite{morigi01a},
\begin{equation}
    \eta=k\beta_*\sqrt{\frac{\hbar}{2m_\textrm{Ca}\omega_*}}.
\end{equation}
Here, $k$ is the laser k-vector and $\beta_*$ is a geometrical factor which is equal to one for the case of a single ion and is equal to\cite{morigi01a}
\begin{equation}
    \beta_-^{Ca}=\sqrt{\frac{\mu}{2}}\frac{1}{\sqrt{\sqrt{1+\mu^2-\mu}\left(\sqrt{1+\mu^2-\mu}+1-\mu\right)}}
\end{equation}
for the case of a two-ion crystal.

\subsection{Simulating the N$_2^+$ ion Rabi sideband signal using a single atomic Ca$^+$ ion}
\label{app:sim}

To simulate the molecular signal using a single atomic ion, we follow these steps:
\begin{enumerate}
    \item Determine the ODF strength, $F_\textrm{ODF,2}$, for a specific molecular quantum state and ODF laser wavelength according to Eq. (\ref{eq:ODF}).
    \item Determine the N$_2^+$-Ca$^+$ two-ion crystal energy, $E_2\left(t_\textrm{ODF}\right)$, after the application of the ODF for time $t_\textrm{ODF}$ according to Eq. (\ref{eq:Etwo}) by numerically solving Eqs. (\ref{eq:eomtwo}).
    \item Determine the Rabi sideband signal for the two-ion crystal according to Eq. (\ref{eq:Rabi}) with $
    \bar{n}_2=\frac{E_2\left(t_\textrm{ODF}\right)}{\hbar\omega_-}$ and $\eta_2=k\beta_-^{Ca}\sqrt{\frac{\hbar}{2m_\textrm{Ca}\omega_-}}$.
\end{enumerate}

At this point, we have determined the expected theoretical molecular signal as can be seen, e.g., in Fig. \ref{fig:HideCa} of the main text. To simulate this signal with a single atomic ion, we need to determine the ODF strength, $F_\textrm{ODF,1}$, which we need to apply on a single ion to get the same signal. This ODF strength is not equal to the ODF strength of the two-ion crystal, $F_\textrm{ODF,2}$, due to the different mechanical configuration of a two-ion crystal and a single ion. Thus: 
\begin{enumerate}
    \item[4.] Find the mean Fock-state coherent distribution, $\bar{n}_1$, which best fits the Rabi sideband molecular signal derived in step 3 using Eq. (\ref{eq:Rabi}) for a single ion with $\eta_1=k\sqrt{\frac{\hbar}{2m_\textrm{Ca}\omega_t}}$.
    \item[5.] Determine the ODF strength, $F_\textrm{ODF,1}$, required to achieve the energy, $E_1=\bar{n}_1\hbar\omega_t$, after application of the force for a duration $t_\textrm{ODF}$ by solving Eq. (\ref{eq:eomsingle}) and using Eq. (\ref{eq:Eone}) for the ion's energy.
    \item[6.] To apply $F_\textrm{ODF,1}$ on a single ion, one can either determine the peak intensity of the laser beams required or measure directly the ac-Stark shift on a single ion from which the ODF strength can be calculated using Eq. (\ref{eq:ODF}).
\end{enumerate}

\begin{figure}
	\centering
	\includegraphics[width=0.5\linewidth,trim={7cm 2cm 7cm 3cm},clip]{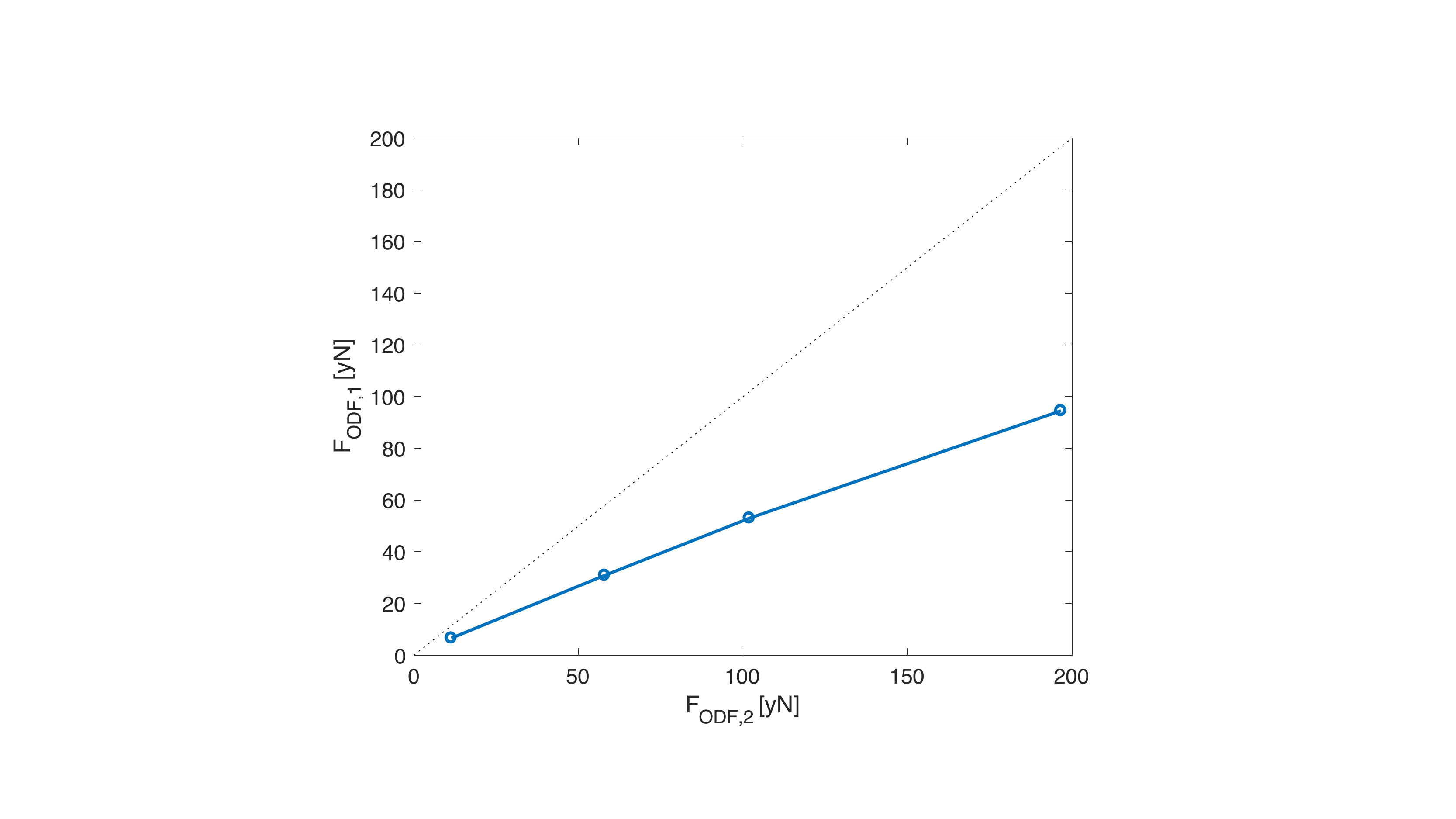}\\
	\caption{The optical-dipole force (ODF) strength in yN=$10^{-24}$ N for a two-ion N$_2^+$-Ca$^+$ crystal, $F_\textrm{ODF,2}$, and single Ca$^+$ ion, $F_\textrm{ODF,1}$, which give rise to the same Rabi sideband signal. The dashed line indicates the non-physical case of $F_\textrm{ODF,2}$=F$_\textrm{ODF,1}$}
	\label{fig:ESI_ODF}
\end{figure}

The results of our ODF strength conversion for the molecular states given in Fig. \ref{fig:N2Qsim} of the main text are summarized in Fig. \ref{fig:ESI_ODF}.  

\subsection{Scattering rate due to the optical-dipole-force laser beam}
\label{app:scatt}

To estimate the error due to off-resonant photon scattering on \Nion{} by the optical lattice, we need to determine the relevant scattering rate. The scattering rate of level $\ket{i}$  due to excitation to level $\ket{j}$ is given by\cite{SteckQuantumAtomOptics}:
\begin{equation}
    \Gamma_{sc}^{(i,j)} = \Gamma^{(j)} \rho_{jj}^{(i)}.
\end{equation}
Here, $\Gamma^{(j)}$ is the natural line width, \textit{i.e.}, the radiative decay rate, and $\rho_{jj}^{(i)}$ is the diagonal density matrix element (the population) of the upper state $\ket{j}$. For large detunings, \textit{i.e.} ${\Delta_{j}^{(i)}} \gg \Omega_j^{(i)},\Gamma^{(j)}$, the upper state population is given by:
\begin{equation}
    \rho_{jj}^{(i)} = \frac{1}{2\epsilon_0\hbar^2 c} I\left(\textbf{x},t\right)
    \frac{\left|\left\langle j \left| \bm{\mu} \right| i\right\rangle \right|^2}{\left(\Delta_{j}^{(i)}\right)^2}.
\end{equation}

The spontaneous decay rate of the upper state for a multilevel system is given by\cite{Grimm1999}:
\begin{equation}
    \Gamma^{(j)} = \frac{1}{3\pi \epsilon_0\hbar c^3} \sum_k \left(\omega_k^{(j)}\right)^3 \left| \left\langle k \left| \mu_{red} \right| j\right\rangle\right|^2 = 1/\tau_j.
\end{equation}
Here, the sum runs over all possible states $\ket{k}$ to which the excited state $\ket{j}$ can decay, $\omega_k^{(j)}$ is the transition frequency, $\left| \left\langle k \left| \mu_{red} \right| j\right\rangle\right|$ is the reduced dipole matrix element of the transition $\ket{j}\rightarrow\ket{k}$ and $\tau_j$ is the life time of state $\ket{j}$. In this treatment, the decay rate is summed over all possible polarizations.

For large detunings, we also need to consider the case of several excited levels $\ket{j}$. On the other hand, the excited-state populations are small for large detunings, \textit{i.e.}, $\rho_{jj}^{(i)} \approx 0$. Therefore, we can treat the different possible transitions to states $\ket{j}$ as independent and calculate the photon scattering rate of state $\ket{i}$ as:
\begin{equation}
    \Gamma_{sc}^{(i)} = \sum_j \Gamma_{sc}^{(i,j)}.
\end{equation}
With this, we can estimate the probability to scatter at least one photon within an exposure time $\tau$ as:
\begin{equation}
    P_{sc} = 1 - e^{-\tau \Gamma_{sc}^{(i)}}.
\end{equation}

For the \Nion{} ion and excitation on the $(2,0)$ vibrational band of the \AX{} transition, we can estimate the upper state decay rate, equal for all rotational states, as:
\begin{equation}
    \Gamma^{(A,v = 2)} \approx \sum_{v"} A_{vib}^{(2,v")}
\end{equation}
Here, $A^{(2,v")}_{vib}$ are the vibronic Einstein A-coefficients for the upper state $v' = 2$. With the tabulated values for $v" = 0 -5$ given in Ref.\cite{Langhoff1987}, we calculated a decay rate for the $A^2\Pi_u, (v = 2)$ state of $\Gamma^{(A,v' = 2)} \approx 85 \ \textrm{kHz}$. 

For the rovibrational ground state of \Nion{} with the $R_{11}(1/2)$, $m_{J"} = m_{J'} = -1/2$ $\pi$-transition as the only possible transition and a beam with $I = 2P/ \pi w_0^2 = 10\cdot10^6\  \textrm{W}/\textrm{m}^2$ at $786.7\ \textrm{nm}$, we obtain a scattering rate of $\Gamma_{sc} \approx 300\ \mu\textrm{Hz}$ and thus within $\tau = 10\ \textrm{s}$ a scattering probability of $P_{sc} \approx 0.3\%$. 

\end{document}